\documentclass[aps, prd, 10pt, twocolumn, superscriptaddress,noshowpacs, preprintnumbers, longbibliography, nofootinbib,bibnotes,hyperref,floatfix]{revtex4-2}

\usepackage{graphicx}
\usepackage{dcolumn}
\usepackage{bm}

\usepackage{amsmath}
\usepackage{array,multirow,graphicx}
\usepackage{ulem}

\usepackage{aas_macros} 

\usepackage{color}
\usepackage{appendix}
\definecolor{cadmiumgreen}{rgb}{0.0, 0.42, 0.24}
\definecolor{orange}{RGB}{255,127,0}

\usepackage{tikz,xcolor,hyperref}

\definecolor{lime}{HTML}{A6CE39}
\DeclareRobustCommand{\orcidicon}{\hspace{-1mm}
	\begin{tikzpicture}
	\draw[lime, fill=lime] (0,0) 
	circle [radius=0.16] 
	node[white] {{\fontfamily{qag}\selectfont \tiny \,ID}};
	\draw[white, fill=white] (-0.0525,0.095) 
	circle [radius=0.007];
	\end{tikzpicture}
	\hspace{-3mm}
}

\foreach \x in {A, ..., Z}{\expandafter\xdef\csname orcid\x\endcsname{\noexpand\href{https://orcid.org/\csname orcidauthor\x\endcsname}
			{\noexpand\orcidicon}}
}

\begin{document}
\title{Multi-messenger observations of core-collapse supernovae: Exploiting the standing accretion shock instability}

\author{Marco Drago\orcidM} 
\affiliation{Universit{\`a} di Roma La Sapienza, Department of Physics, 00133 Roma, Italy}\affiliation{INFN, Sezione di Roma, 00133 Roma, Italy}

\author{Haakon Andresen\orcidA} 
\affiliation{The Oskar Klein Centre, Department of Astronomy, AlbaNova, SE-106 91 Stockholm, Sweden}

\author{Irene Di Palma\orcidI} 
\affiliation{Universit{\`a} di Roma La Sapienza, Department of Physics, 00133 Roma, Italy}\affiliation{INFN, Sezione di Roma, 00133 Roma, Italy}

\author{Irene Tamborra\orcidB{}}
\affiliation{Niels Bohr International Academy \& DARK, Niels Bohr Institute,\\University of Copenhagen, Blegdamsvej 17, 2100 Copenhagen, Denmark}

\author{Alejandro~Torres-Forn\'e\orcidT{}}
\affiliation{Departamento de Astronom\'ia y Astrof\'isica, Universitat de Val\`encia, E-46100 Burjassot, Val\`encia, Spain}
\affiliation{Observatori Astron\`omic, Universitat de Val\`encia, E-46980 Paterna, Val\`encia, Spain}

\date{\today}
\begin{abstract}

The gravitational wave (GW) and neutrino signals from core-collapse supernovae (CCSNe) are expected to carry pronounced imprints of the standing accretion shock instability (SASI). 
We investigate whether the correlation between the SASI signatures in the GW and neutrino signals could be exploited to enhance the detection efficiency of GWs. We rely on a benchmark full-scale three-dimensional CCSN simulation with zero-age main sequence mass of $27\ M_\odot$. Two search strategies are explored: 
1.~the inference of the SASI frequency range and/or time window from the neutrino event rate detectable at the IceCube Neutrino Observatory; 2.~the use of the neutrino event rate to build a matched filter template.
We find that incorporating information from the SASI modulations of the IceCube neutrino event rate can increase the detection efficiency compared to standard GW excess energy searches up to $30\%$ for nearby CCSNe. However, we do not find significant improvements in the overall GW detection efficiency for CCSNe more distant than $1.5$~kpc.
We demonstrate that the matched filter approach performs better than the unmodeled search method, which relies on a frequency bandpass inferred from the neutrino signal. The improved detection efficiency provided by our matched filter method calls for additional work
 to outline the best strategy for the first GW detection from CCSNe.

\end{abstract}

\keywords{Suggested keywords}
\maketitle

\section{Introduction}
\label{sec_intro}

The collapse of massive stars (with mass between $8~{\rm M}_\odot$ and $100~{M}_\odot$) leads to copious production of neutrinos as well as gravitational waves (GWs). By the end of their lives, these stars have developed an iron core, whose mass is close to the Chandrasekhar mass limit. When the pressure of relativistic degenerate electrons can no longer balance gravity, the core collapses.
The collapse continues until the inner core reaches nuclear densities, the repulsive forces between nuclei stabilizes the
inner core which forms a proto-neutron star (PNS). The, still in-falling, outer core rebounds off the newly formed PNS
and a shock wave is launched. The shock travels through the in-falling material, but energy loss from 
photo-dissociation of iron nuclei and neutrino cooling causes the shock to halt after reaching a radius of
around $150$~km \cite{Janka:2006fh,Mezzacappa:2020oyq,Burrows:2020qrp}.
According to the neutrino-driven mechanism, the stalled shock is revitalized through neutrinos
streaming away from the PNS. 
Neutrinos deposit energy in the medium behind the shock, enabling the core-collapse supernova (CCSN) explosion \cite{Janka:2000bt}.
The shock wave propagates through the collapsing star, unbinding and ejecting the outer layers. 
If the neutrino energy deposition is insufficient then the shock wave is not revived and the in-falling material continues to accrete through the shock wave. As a result, the mass of the central compact object
continues to increase and eventually leading to black hole formation \cite{OConnor:2012bsj,Walk:2019miz}.
The time between the recoil of the core and the revitalization of the shock/black hole formation
is referred to as the post-bounce phase. It is expected that most of the GW emission is emitted during
this time.
During the accretion phase, large-scale convective motions behind the shock and the standing accretion shock instability (SASI) \cite{Blondin:2002sm,Blondin:2006yw,Scheck:2007gw} can enhance the energy transfer to the medium, imprinting characteristic signatures on the neutrino and GW signals \cite{Lund:2010kh,Lund:2012vm,tamborra_13,tamborra_14b, Kuroda:2016bjd,Mezzacappa:2020lsn,Walk:2022eld,Walk:2019miz,Walk:2019ier,Takiwaki:2017tpe,Takiwaki:2021dve}.

 The numerical simulation of the core-collapse scenario is an arduous computational task \cite{Mezzacappa:2020oyq}. In addition, the physics of the core collapse is still affected by several uncertainties \cite{Mezzacappa:2020oyq,Burrows:2020qrp,Tamborra:2020cul,Richers:2022zug,Duan:2010bg,Oertel:2016bki,Burrows:2016ohd}. 
Given their complex nature, hydrodynamical instabilities manifest differently in the existing set of CCSN simulations, with related implications on the expected neutrino and GW signals.
 Unlike the compact binary merger scenario, there is no univocal correlation between the properties of the progenitor star and the resulting GW waveforms. The absence of correlation and  our lack of conceptual understanding of some of the aspects linked to the development of hydrodynamical instabilities present a challenge in accurately predicting the GW waveforms from CCSN. 

Recent multi-dimensional simulations of CCSNe exhibit distinctive GW signal features that are correlated with the post-bounce hydrodynamic activity, such as oscillations of the PNS and the SASI
 \cite{Marek:2008qi, Mueller:2012sv, Kuroda:2016bjd, Andresen:2016pdt, Powell:2018isq, Radice:2018usf, Andresen:2020jci, Mezzacappa:2020lsn, Vartanyan:2020nmt,Andersen:2021vzo, 
 Mezzacappa:2022xmf, Vartanyan:2023sxm}.
The GW waveforms
 display stochastic behavior, which arises from the complex hydrodynamical physics that occur during the post-bounce phase.
However, simulations from all groups show common features in the time-frequency domain. The most noticeable is characterized by a frequency that rises from roughly $100$~Hz to a few kHz (at most) and lasts for $0.5$--$1$~s. This feature is interpreted as a gravity mode (g-mode) that is continuously excited in the PNS \cite{Murphy:2009, Mueller:2012sv, Yakunin:2015wra, Cerda-Duran:2013swa, Kuroda:2016bjd, Andresen:2016pdt, Torres-Forne:2017xhv, Torres-Forne:2018nzj}.
Additionally, Refs.~\cite{Cerda-Duran:2013swa, Kuroda:2016bjd,Andresen:2016pdt} have related the presence of a low-frequency ($\approx 100$~Hz) component in the GW signal characteristic frequency of SASI. Therefore, it is possible to detect GWs at different frequencies, according to whether they carry imprints from PNS or SASI activity. More recently, it has been pointed out that GW emission might be expected from the cocoon inflated by the jet in the case of collapsars~\cite{Gottlieb:2022qow}; in this scenario, one should expect GW emission in the frequency range of $10$--$600$~Hz over the jet timescale. 

Even though the range of frequencies of the GW emission falls in the band of detection of current GW detectors, 
it is expected that for slowly/non-rotating CCSNe
can be detected at  distances up to $\sim 5$~kpc \cite{Szczepanczyk:2021bka,Abdikamalov:2020jzn}. At this distance CCSNs have an event rate of about $2$--$3$ per century in the Galaxy \cite{Rozwadowska:2020nab}, which challenges the detection prospects for Advanced LIGO \cite{LIGOScientific:2014pky}, Advanced Virgo \cite{VIRGO:2014yos}, and KAGRA \cite{Aso:2013eba}.
The predicted event rate of third generation detectors, i.e.~the Einstein Telescope \cite{Punturo:2010zza} or Cosmic Explorer \cite{Reitze:2019iox}, increases to $\sim0.5~ \rm{yr}^{-1}$ \cite{Ando:2005ka}, while the predicted detection horizon can extend up to $2$--$4$~Mpc \cite{Sathyaprakash:2012jk} in the case of rapidly rotating progenitors.
If the progenitor core exhibits very rapid rotation, observed only in about $1\%$ of the electromagnetically observed events \cite{Li:2010kc, Chapman:2007vp}, the magnetic fields can transfer the rotational kinetic energy of the PNS to the shock. This mechanism produces an explosion that will lead to an enhanced GW signature that can be detectable within a distance of $50$~kpc, and in certain extreme models, extend up to $5$--$30$ Mpc \cite{Gossan:2015xda, LIGOScientific:2019ryq}. Alternatively, the GW emission from the cocoon could be detectable up to $200$~Mpc by existing interferometers~\cite{Gottlieb:2022qow}.

Given the computational costs of CCSN simulations as well as the stochastic nature of the GW signals and the considerable number of physical parameters involved in the explosion mechanism, it is not possible to generate the dense bank of templates required by the common data analysis techniques based on template matching.
To date, the data analysis methods for CCSNe primarily focus on reconstructing the GW strain amplitude, without assuming a specific signal model \cite{klimenko2008coherent, drago2021coherent}. 
Other works have applied the Principal Component Analysis, Bayesian model selection, and dictionary learning to reconstruct the CCSN waveform and infer the explosion mechanism \cite{Summerscales:2007xq, Heng:2009zz, Rover:2009ia, Powell:2018csz, Roma:2019kcd,Saiz-Perez:2021bce}.
Additionally, CCSN parameter inference has been attempted in the literature \cite{Bizouard:2020sws,Bruel:2023iye, Torres-Forne:2019zwz}. 

In this paper, we investigate complementary methods with respect to the ones exploiting PNS oscillations. We explore the possibility of using the information linked to the SASI signatures present in the detectable neutrino event rate to inform the search for GWs. 
This paper is organized as follows. Section \ref{sec_gwnusignals} describes the main characteristics of the neutrino and GW signals from our benchmark CCSN model. Section \ref{sec_detectors} briefly describes the main characteristics of neutrino and GW detectors.  We introduce the algorithms adopted in this work for GW searches in Sec.~\ref{sec_method}, our results as well as a comparison on the detection efficiencies are instead presented in Sec.~\ref{sec_results}. Finally, a discussion on our findings is provided in Sec.~\ref{sec_discussion}. In Sec.~\ref{sec_conclusions} we conclude. 


\section{Gravitational wave and neutrino signals}
\label{sec_gwnusignals}

The neutrino and gravitational wave signals adopted in this work have been extracted from
our benchmark CCSN simulation: the model s27 presented in Ref.~\cite{Hanke:2013ena}.
The stellar progenitor of the model s27 is a non-rotating star of solar metallicity
with zero-age main sequence mass of $27~M_\odot$~\citep{woosley_02}.
The CCSN
simulation in three spatial dimensions with sophisticated three-flavor neutrino transport has been carried 
out with \textsc{Prometheus-Vertex}. 
The \textsc{Prometheus-Vertex} code~\cite{rampp_02,buras_06a} 
is split into two main modules. The
hydrodynamics calculations are performed by \textsc{Prometheus}~\cite{mueller_91,fryxell_91}.
The monopole approximation,
with a pseudo-relativistic potential~\cite{marek_06},
is used to treat self-gravity.
The module \textsc{Vertex}~\cite{rampp_02} handles neutrino transport by
solving the energy-dependent two-moment equations for three neutrino species,
electron neutrinos ($\nu_e$), electron antineutrinos ($\bar{\nu}_e$),
and a third species ($\nu_x$) representing muon or tauon flavors. The neutrino radiation transport is calculated in the
``ray-by-ray-plus'' approximation~\cite{buras_06a}.
The high-density equation of state of Lattimer and Swesty~\cite{lattimer_91} with
 nuclear incompressibility of $K=220 \,\mathrm{MeV}$ was used for the s27 simulation. 
 We refer the interested reader to Ref.~\cite{Hanke:2013ena} for a more detailed description of the simulation setup.
 
 The model s27 has been simulated until $550$~ms post bounce and 
 it exhibits two periods of strong SASI activity with a period of convective overturn in between. 
 The first SASI phase takes place between $120$ and $260 \, \mathrm{ms}$ 
after the start of the simulation
and the second phase starts around $410 \, \mathrm{ms}$ and lasts until the end of the simulation. 
 Detailed analyses of the neutrino emission properties and the gravitational-wave signal are reported in Refs.~\cite{tamborra_13,tamborra_14b,Andresen:2016pdt}. 

An observer 
at $10$ kpc from the source is expected to observe a GW
strain of $\mathcal{O}(10^{-23})$. 
The GW signal of the model s27 is dominated by 
two main emission components \cite{Andresen:2016pdt}: 
oscillations of the PNS lead to an emission above 
$300$~Hz with a central frequency that
grows approximately with time; 
SASI is responsible for GW emission between $75$ and $250$~Hz, during the two SASI periods.
SASI also induces large-amplitude periodic modulations in the neutrino signal. Such modulations remain clearly visible even for the observable neutrino flux which is given by an hemispheric average performed to include flux projection effects in the observer direction~\cite{tamborra_14b}.
 The detectable central frequency characterizing the SASI modulations for the model s27 is approximately between $50$ and $120$~Hz, peaking around $85$~Hz~\citep{tamborra_13}. 
Note that the timescale of SASI, its magnitude, and the number of SASI episodes depend on the
properties of the CCSN simulation and can lead to variations that are in turn reflected in the frequency of the modulation imprinted on the
neutrino and GW signals, see e.g.~Refs.~\cite{tamborra_13,tamborra_14b,Walk:2019miz,Walk:2019ier,Walk:2018gaw}. 

The SASI modulations of the neutrino signal are present for all neutrino flavors, although they have a smaller amplitude for the non-electron flavors as pointed out in Ref.~\cite{tamborra_14b}, see e.g.~their Fig.~7 and related discussion. Because of flavor conversion occurring in the stellar core~\cite{Mirizzi:2015eza,Tamborra:2020cul}, the neutrino signal observed at Earth from the next CCSN explosion will likely be a mixture between the electron and non-electron neutrino flavors. Given our preliminary understanding of flavor conversion in CCSNe~\cite{Tamborra:2020cul,Chakraborty:2016yeg,Duan:2010bg}, we refrain from assuming any specific flavor conversion scenario. Instead we consider two extreme cases for the neutrino signal detected at Earth through inverse beta decay at the IceCube Neutrino Observatory: absence of flavor conversion (in this case, we would detect the unoscillated $\bar\nu_e$ signal) and full flavor conversion (in this case, we would detect the unoscillated $\bar\nu_x$ signal). 

It is worth noticing that SASI manifests in the neutrino and GW signals as a direction-dependent phenomenon. If SASI  mostly develops along a plane (like it initially happens for our s27 model~\cite{Hanke:2013ena,tamborra_14b}), then  the neutrino signal has modulations that are stronger for an observer located along the SASI plane and weaker for an observer located perpendicularly to the SASI plane~\cite{tamborra_13}. 
On the other hand, one expects the GW signal to be maximally modulated for an observer
situated perpendicularly to the observer seeing the strongest SASI modulations in the
neutrino signal. 
Yet, SASI is a global instability which affects
the whole simulation volume; furthermore, the neutrino and GW observable signals depend on volume 
integrated quantities (projected onto any given observer direction). 
Consequently, we observe
extended regions where the correlation between the neutrino and GW signals is better or worse, compared
to stronger (anti-)correlations seen along the directions where SASI is more prominent. 

\begin{figure*}
 \centering
 \includegraphics[width=0.99\textwidth]{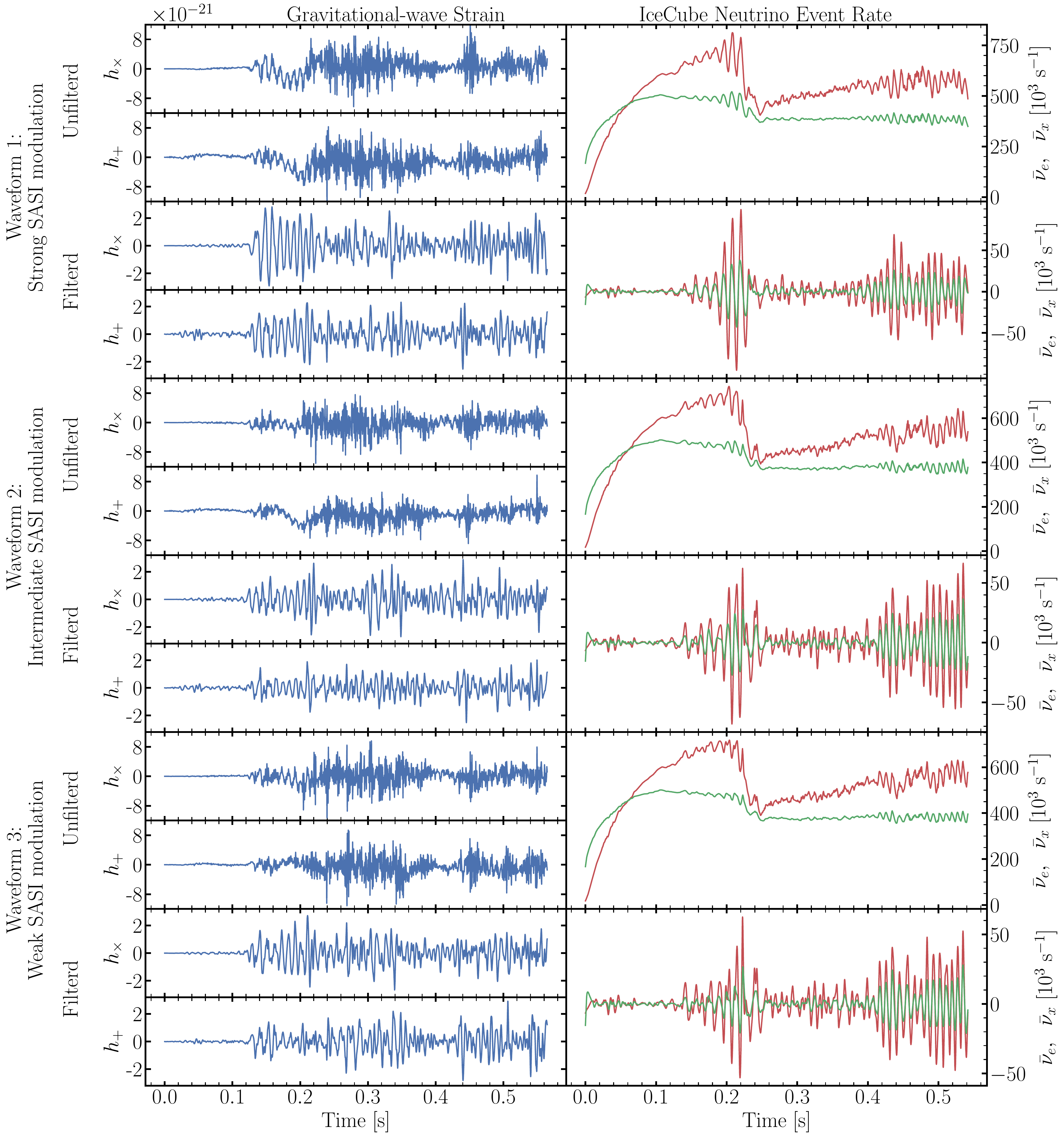}
 \caption{Gravitational wave strain (left column, in blue) and IceCube event rate for $\bar\nu_e$ (no flavor conversion, in red) and $\bar\nu_x$ (full flavor conversion, in green) (right column) for our benchmark CCSN model, assuming a CCSN at $10$~kpc. The observer directions are selected for strong (top plots), intermediate (middle plots) and weak (bottom plots) neutrino SASI modulations, respectively. The sub-panels in the
 left plots show the cross (top) and plus
 (bottom) polarization modes of the GW signal. 
 The bottom panels, labeled ``Filtered,'' show the
 GW (left) and neutrino signals (right) after low-pass and high-pass band filters have been
 applied. The filters remove any part of the signal below $50$~Hz and above $300$~Hz. The filtered GW and neutrino signals show a certain degree of correlation in their modulations.} 
 \label{fig:filterdunfilterd}
\end{figure*}

Figure~\ref{fig:filterdunfilterd} shows an example of the GW strain and neutrino event rate expected in the IceCube Neutrino Observatory for a CCSN located at $10$~kpc (note that the CCSN distance will be tailored for illustrative purposes in the following sections) and for the two flavor conversion scenarios. The selected observer directions correspond to the one along which the neutrino event rate has the strongest (waveform 1), intermediate (waveform 2) and weakest (waveform 3) SASI modulations, respectively, as from Refs.~\cite{tamborra_13,tamborra_14b}. After computing the neutrino energy distributions from the model s27 in the form of Gamma distributions~\cite{Keil:2002in,Tamborra:2012ac} and taking into account observer projection and limb-darkening effects~\cite{tamborra_14b}, the IceCube event rate has been implemented following Ref.~\cite{tamborra_13}--see also Sec.~\ref{sec_detectors}.
For each observer direction, the top panels of Fig.~\ref{fig:filterdunfilterd} show the unfiltered signals, obtained from the model s27.
The bottom panels show the same GW (left) and neutrino (right) signals after filtering out 
frequencies outside the band $\Delta f \in [50, 300]$~Hz.
The bottom panels of Fig.~\ref{fig:filterdunfilterd} illustrate that 
SASI modulations are evident in both the GW and neutrino signals
and that there is a degree of correlation between the modulations of 
these two messengers. 

\section{Neutrino and Gravitational Wave Detectors}\label{sec_detectors}
The largest operating detectors sensitive to CCSN neutrinos are Cherenkov detectors: the IceCube Neutrino Observatory and Super-Kamiokande. Neutrinos are mainly detected through the inverse beta decay channel ($\bar\nu_e p \rightarrow n + e^+$). IceCube is expected to register about $10^6$~events above background for a CCSN at $10$~kpc, while Super-Kamiokande would approximately detect $10^4$ events and it would practically be background free~\cite{Mirizzi:2015eza,Scholberg:2017czd}. 

As shown in Fig.~\ref{fig:filterdunfilterd}, we consider the neutrino signal expected in IceCube given its largest statistics and in order to optimize the detection prospects of GWs that are already known to be limited to CCSNe occurring within the Galaxy. However, one should keep in mind that the upcoming Hyper-Kamiokande~\cite{Abe:2011ts} will provide a background-free signal of roughly $1/3$ the IceCube rate; hence Hyper-Kamiokande may achieve better signal statistics than IceCube at larger CCSN distances for stellar collapses exhibiting very strong and long-lasting SASI activity. The IceCube event rate is implemented following Refs.~\cite{tamborra_13,tamborra_14b,IceCube:2011cwc}. 

Gravitational wave detectors entered the Advanced Era in 2015 with the beginning of the first scientific run. The two detectors developed by the LIGO Collaboration \cite{LIGOScientific:2014pky}, located in Livingston and Hanford, routinely collect data. Advanced Virgo \cite{VIRGO:2014yos} joined at the end of the second run, in August 2017, contributing to the multi-messenger detection of the binary neutron star merger event GW170817 on August 17th. In Japan, the KAGRA Collaboration built another detector \cite{Aso:2013eba}, which already joined the third observing run, but it is not considered in this analysis. 

In the following, 
we consider the IceCube neutrino event rate and a GW-detector network composed of three detectors: the two LIGO detectors and Virgo. The random shot noise for the neutrino detector follows Poisson statistics and does not depend on the CCSN distance, while the neutrino event rate scales as the inverse square of the distance~\cite{tamborra_13}. For the data generation of GW detectors, we adopt the design advanced spectral sensitivities (named as O5 in Fig. 1 of ~\cite{KAGRA:2013rdx}). The standard way to generate noise for GW detectors is to simulate Gaussian noise with null mean and sigma equal to one. These data are then transformed in the Fourier domain, where they show a whiten spectrum. To resemble the desired GW data, we multiply this frequency series for the selected sensitivity and return in the time domain to obtain a series of amplitude similar to the expected detector data.

\section{Search methods for gravitational waves}
\label{sec_method}
In this section, we summarize the standard LIGO-Virgo unmodeled search method, 
in particular the one used for CCSNe.
We then outline the algorithms that we have developed for the search of GWs by taking advantage of the signal modulations due to SASI, matching the timing and frequency information from the neutrino event rate with the ones of the GW signal.

\subsection{Unmodeled gravitational wave searches}
The GWs detected so far originate from the coalescence of compact objects in a binary system. In this case, the GW waveforms are well known and can be characterized by a set of templates, which resemble the detectable signal. Optimal algorithms for this search adopt a matched filter approach, i.e.~find the optimal match among one of the templates and the detector data \cite{LIGOScientific:2019hgc}.

For CCSNe, the GW signal is strongly dependent on the properties of the collapsing star and the eventual presence of hydrodynamical instabilities.
The large variety of features potentially present in the GW signal renders the matched filter approach sub-optimal.
The alternative adopted solution is the so-called \textit{un-modelled} search method: looking for signals of uncertain morphology through a statistically significant excess of power in the detector data in the time-frequency plane (\textit{excess-power method}).

The excess power algorithm adopted by the LIGO-Virgo Collaboration in the case of CCSN alerts from electromagnetic triggers \cite{Abbott:2019pxc} is the so-called coherent WaveBurst\footnote{https://doi.org/10.5281/zenodo.4419902} (cWB) \cite{klimenko2008coherent, Klimenko:2015ypf, drago2021coherent}. 
cWB takes into account a minimal approach for the search of noise excess in the {time-frequency domain, obtained from data } after applying a Wilson-Daubechies-Meyer wavelet transform \cite{Klimenko:2004qh}.
A combined likelihood across the data of the three interferometers allows to assess the properties of the GW candidate events. 

Given that unmodeled algorithms are suited to search for any kind of signal, they are also plagued by transient noise, which affects the distribution of false alarms. Internal parameters, called regulators, significantly reduce the noise excess, especially in the case of two aligned detectors like the two LIGO ones \cite{Klimenko:2015ypf}. 
This search method has been used in past analyses for GW emission from CCSNe focused on the use of the two LIGO detectors \cite{Abbott:2019pxc}. Even if a fine-tuning of the algorithm would be desiderable once the Virgo detector is included, such a task is out of the scope for this paper. Hence, we rely on the 
same analysis configuration adopted in the LIGO-Virgo searches \cite{Abbott:2019pxc} as a benchmark method to be compared with the GW search methods explored in this paper and introduced in the next sections (Secs.~\ref{sec:GWnu_method} and \ref{sec:nutrigger}).

A recent review on GW searches from CCSNe with the cWB method is reported in Ref.~\cite{Szczepanczyk:2021bka}, where different detector sensitivities and detector networks are explored. Recent attempts also aim to exploit machine learning techniques to improve the detection capability of CCSN searches \cite{Cuoco:2020ogp, Iess:2020yqj, Cavaglia:2020qzp, Astone:2018uge, Edwards:2020hmd, Chan:2019fuz, LopezPortilla:2020odz}; however, since they are not officially used by the LIGO-Virgo collaboration \cite{Abbott:2019pxc} yet, we do not consider them for this work.

\subsection{Gravitational wave searches with neutrino matched filter}
\label{sec:GWnu_method}

In order to investigate the potential of neutrinos in aiding GW searches from CCSNe, building on the correlation in the modulations of the GW and neutrino signals shown in Fig.~\ref{fig:filterdunfilterd}, we have developed a template-based triggered search. Relying on the neutrino event rate detected by IceCube displayed in Fig.~\ref{fig:filterdunfilterd}, we build a template to be used in a matched filter approach. Note that we are interested in exploiting the SASI frequency and the SASI time windows inferred from the neutrino event rate; both of them can be determined with relatively high precision~\cite{tamborra_13,tamborra_14a}.

Since we intend to develop a matched filter technique, we consider one of the standard LIGO-Virgo algorithms: the PyCBC library\footnote{https://doi.org/10.5281/zenodo.596388} \cite{Allen:2004gu,Allen:2005fk,Nitz:2017svb,DalCanton:2014hxh}.
This is a software package developed for the search of coalescing compact binaries, which can additionally measure the astrophysical parameters of detected sources. PyCBC has been used in the analysis of the Advanced LIGO and Virgo observing runs \cite{LIGOScientific:2021djp, LIGOScientific:2020ibl, LIGOScientific:2018mvr}, participating in the first direct detection of GWs \cite{LIGOScientific:2016aoc}.

\begin{figure}
 \centering
 \includegraphics[width=9cm]{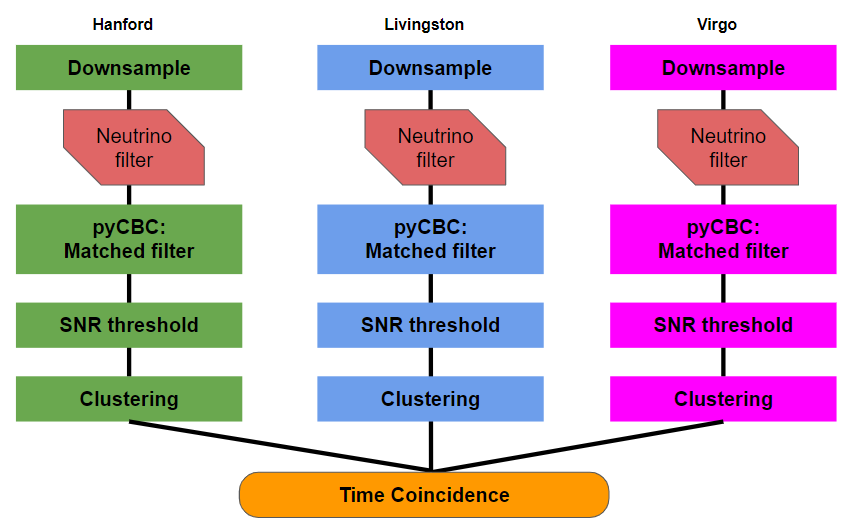}
 \caption{Flowchart of the novel matched filter pipeline proposed in this paper to search for GWs by relying on the SASI signatures detectable in the neutrino signal.}
 \label{fig:flowchart}
\end{figure}

We have created a novel pipeline, mixing the features of the matched filter and unmodeled searches; the underlying assumption is that the SASI signatures in the neutrino event rate should be correlated in time to the GW signal features. This enables us to exploit time-selection and/or band-pass filters in the time/frequency domain.
The flowchart of this novel pipeline is sketched in Fig.~\ref{fig:flowchart}. Its main steps are:
\begin{enumerate}
 \item Downsampling of the original data to a defined sample rate. This reduces the computational load since the GW signal is expected to be below $2000$ Hz while the GW interferometers take data with a sample rate of $16384$~Hz. 
 \item Applying the pyCBC function filter.matched\_filter between the theoretical template and the data stream. The result is a time series which contains the matched SNR.
 \item Selecting samples of the signal time series with an SNR bigger than a selected threshold value ($SNR_{\rm{th}}$).
 \item It is possible that the same template gives high values of SNR in consecutive samples of the time series. To avoid to consider the same event multiple times, we merge consecutive samples within a unique cluster, if their times differ less than a temporal threshold $\Delta t_c$. The clusters are characterized by three parameters that will be used in the next step:
 \begin{itemize}
 \item SNR: maximum SNR in the samples;
 \item {Time}: the GPS time corresponding to the maximum SNR;
 \item {Occupancy (O)}: the number of samples.
 \end{itemize}
 \item If the GPS times characteristics of the clusters coming from different detector analyses differ less than a chosen value $\Delta t_n$, they are grouped within a single event, which is considered as the final trigger of the entire analysis.
\end{enumerate}
 The outcome of this procedure consists of a list of network triggers, which is characterized by the two following quantities:
\begin{equation*}
\begin{matrix}
 SNR_{\rm{net}} = \sqrt{\sum_i SNR^2_i}\ &\text{and} &
 O_{\rm{net}} = \sum_i O_i
\end{matrix}
\end{equation*}
where $i = 1, ..., 3$ in this work since we rely on three GW detectors. The same definition can be applied for any number of detectors.

The performance of the pipeline strictly depends on the algorithm parameters that we have introduced (i.e., the analysis sample rate, the SNR threshold of each sample, the time coincidence between single samples $\Delta t_c$, and the time coincidence among clusters from different detectors $\Delta t_n$). 

\subsection{Gravitational wave searches with neutrino trigger in frequency}
\label{sec:nutrigger}
We also exploit the possibility to feed cWB with additional information from the neutrino event rate to guide the GW search. In fact, we can identify the time periods and frequency bandwidth characteristics of the SASI activity with high precision from the detectable neutrino signal~\cite{tamborra_13,tamborra_14a}.

In this work, we choose to provide to cWB only with the frequency information, and not the temporal one. This is done for two reasons. First, the technical structure of the cWB routine for what concerns the wavelet transform challenges the implementation of the time windows of SASI. The algorithm uses multiple time-frequency domains which differ among them for the resolution of time and frequency pixels. Specifically, this is identified by the level resolution, passing from a level to the next one, the frequency resolution halves and the time resolution doubles. Moreover, we cannot perform too small time selection in a given trigger (like $\mathcal{O}(100$~ms$)$, as visible from Fig.~\ref{fig:filterdunfilterd}), because if this selection is smaller than the time-resolution of the time-frequency pixels of the wavelet transform then it would not work. This because cWB pipeline requires a time-frequency coincidence between the time-frequency pixels from at least different wavelet resolutions.
Second, we simulate GW detector noise which is Gaussian and stationary, hence the noise realization near an injected CCSN signal is statistically compatible with any simulated detector data. The effect can be different in case of real data, where the noise detector can be statistically different for specific time frames and a constraint on time could affect the efficiency. 
We expect that the extraction of the SASI characteristics times from the neutrino event rate might lead to an 
additional improvement on the GW detection sensitivity and leave it to future work.

\section{Results}
\label{sec_results}
In this section, we present our findings on the potential of using the SASI signatures imprinted in the neutrino and GW signals to improve the GW detection efficiency. 
For completeness, we contrast these methods with cWB. 

\subsection{Input for gravitational wave searches}
The time of core bounce can be
reconstructed from the detectable neutrino event rate with an error of a few ms~\cite{Halzen:2009sm,Pagliaroli:2009qy}. By narrowing the search window, determining the 
bounce time allows a GW detection pipeline to search the time with the maximum GW SNR~\cite{Nakamura:2016kkl}.
In this paper, we focus on the time window where the neutrino event rate is affected by SASI modulations.
Since these modulations occur during the accretion phase, when the neutrino event rate is larger than the one soon after the bounce,
we expect excellent determination of the temporal window where SASI takes place through the IceCube event rate. The determination of the SASI frequency range from the IceCube event rate could also be considered excellent, as shown in Ref.~\cite{tamborra_13}. 

The observed GW and neutrino signals depend on the orientation of the observer relative to the CCSN event.
In our analysis, we select three representative observer directions
(see Fig.~\ref{fig:filterdunfilterd}) selected according to the strength of the SASI modulations seen in the neutrino signal.
We adopt the following procedure: we choose the neutrino event rate for an observer located along one of the selected directions as well as the related GW signal introduced in Sec.~\ref{sec_gwnusignals}. We then simulate a uniform CCSN distribution in the sky by relying on a sample of $500$ sky directions chosen randomly from a uniform distribution and simulate the arrival of the two messengers from a generic direction in the sky. Next, we generate interferometer data and inject the GW signal.

 We infer from the neutrino event rate that two SASI periods occur in 
the time windows of $[120, 260]$~ms and $[410, 550]$~ms for our benchmark CCSN.
Using the neutrino event rate, we also infer that the GW frequency due to SASI modulations would be between $50$ and $300$~Hz.
In fact, 
based on theoretical predictions, we expect GW signal produced by SASI modulations to have frequency of the same order of magnitude
to that of the neutrino signal. Note that the selected GW frequency range is also in agreement with the findings of a broad range of hydrodynamical simulations.

We use the SASI information inferred from the neutrino event rate in two different ways:
\begin{enumerate}
 \item build a matched filter technique relying on the detected neutrino event rate. 
 Templates are constructed by applying time and/or frequency filters to the neutrino rate (cf.~Fig.~\ref{fig:filterdunfilterd}). We identify this method with the name Matched Filter (MF);
 \item use cWB to perform an Excess Power (EP) search, restricting the frequency band based on the 
frequency observed in the neutrino signal.
\end{enumerate}

In the case of the MF method, since we have two SASI periods for our benchmark CCSN, we repeat our analysis three times: for the first SASI period (MF($t_1$)), the second SASI period (MF($t_2$)), as well as both SASI periods (MF($t_{1, 2}$)). We then consider the three cases above, adding frequency information (MT($t_i$,$f$), with $i=1$, $2$).
The resulting six MF combinations are summarized in Table \ref{Tab: TimeFreq selections}. The two cWB based EP searches we perform are also listed in Table \ref{Tab: TimeFreq selections}; 
we consider a standard unrestricted cWB search in addition to the frequency restricted search outlined above. The standard 
cWB search constitutes our comparison baseline.

\begin{table}[]
 \centering
 \caption{Time and frequency information extracted from the neutrino event rate affected by SASI modulations (see Fig.~\ref{fig:filterdunfilterd}) and used as a template from the GW search. The first column reports the method identifier. The second column indicates the measured time interval during which SASI occurs. The third column lists the frequency band pass.}
 \begin{tabular}{ccc}
 \hline \hline
 Method & Time [ms] & Frequency [Hz] \\
 \hline \hline
 MF($t_1$) & $[120,260]$ & NA\\
 MF($t_2$) & $[410, 550]$ & NA\\
 MF($t_{1, 2}$) & $[120, 260]$--$[410,550]$ & NA\\
 MF($t_1$, $f$) & $[120, 260]$ & $[50, 300]$ \\
 MF($t_2$, $f$) & $[410, 550]$ & $[50, 300]$ \\
 MF($t_{1, 2}$, $f$) & $[120, 260]$--$[410, 550]$ & $[50, 300]$\\
 \hline
 EP & NA & NA\\
 EP($f$) & NA & $[50, 300]$\\
  \hline \hline
 \end{tabular}
 \label{Tab: TimeFreq selections}
\end{table}

Applying the search methods introduced in Sec.~\ref{sec_method}, we calculate the 
detection efficiency at a given \textit{False Alarm Rate} (FAR). 
The detection efficiency is calculated as the sum fraction of total of injected CCSN signals with a FAR smaller than the chosen threshold.
FAR defines the probability that a trigger in the GW detector network may be due to the detector noise. The FAR is assessed following the standard LIGO-Virgo time-shift procedure~\cite{Was:2009vh}: data of different GW detectors are time shifted with a shift considerably bigger than the time of flight among detectors\footnote{The maximum time of flight among the LIGO-Virgo detectors is $27$~ms.}. By doing so, we make sure that coincidence triggers cannot be due to GWs but are random noise in coincidence among the different detectors. The list of coincident 
GW triggers of the shifted data (i.e.~background) resemble the contribution of noise excess giving rise to coincidences in the non-shifted data (i.e.~on source). Repeating the time-shifting procedure multiple times, we obtain the distribution of false alarms characterized by the two parameters introduced in Sec.~\ref{sec_method}: $SNR_{\rm{net}}$ and $O_{\rm{net}}$.  
Given a trigger from the on-source analysis with some values of $SNR_{\rm{net}}$ and $O_{\rm{net}}$, its FAR is defined as the number of triggers from the background analysis that have a $SNR_{\rm{net}}$ and $O_{\rm{net}}$ greater than those of the trigger divided by the total background lifetime.
In the following, we assume that the CCSN signal is detected if its FAR is less than $1$~event per $1$~year.

Each analysis is implemented as a time-triggered search: we search for GWs near the beginning of the SASI window determined through the IceCube event rate. 
Since the detector noise is not stationary, the background for a specific trigger 
is usually estimated from detector data temporally adjacent to the trigger.
However, since we consider simulated noise, which is Gaussian and stationary in time~\cite{Torres:2014zoa}, all the background realizations for each injection are expected to be statistically compatible. To reduce the computational costs we  construct three background realizations (one for each waveform of Fig. \ref{fig:filterdunfilterd}) and use them as benchmark references for all the injections.

First, we need to tune the parameters of the new pipeline. These parameters are: the analysis sample rate, the SNR threshold of each sample ($SNR_{thr}$), the time coincidence between single samples $\Delta t_c$, and the time coincidence among clusters from different detectors $\Delta t_n$. We tune the pipeline considering our benchmark CCSN to be at $0.1$~kpc, the distance is chosen so that the GW signal has 
an SNR ($SNR>20$) that in the entire procedure remains distinguishable from noise and performances are compatible between the different scenarios of Table \ref{Tab: TimeFreq selections}. The parameter set we identified is summarized in Tab.~\ref{Tab. tuning}. 
\begin{table}[h]
 \caption{Running parameters of the MF pipeline which have been selected after a tuning at distance working point of 0.1 kpc. We note that the performances of MF method depend on the choice of the running parameters and therefore we have chosen them such to be conservative.}\label{Tab. tuning}
 \centering
 \begin{tabular}{cccc}
 \hline
 \hline
 Sample rate [Hz] & $SNR_{\rm{thr}}$ & $\Delta t_c$ [ms] & $\Delta t_n$ [ms] \\
 \hline
 \hline
 $4096$ & $3.5$ & $250$ & $300$\\ 
 \hline
 \hline
 \end{tabular}
\end{table}

In the next sections we will compare the efficiencies of all cases in Table~\ref{Tab: TimeFreq selections} for a different distance than the tuning one to verify we have not over-tuned our algorithm for the distance of $0.1$~kpc but the chosen running parameters are optimal also at different distances. We have chosen a distance of $0.5$~kpc away, which is close enough so the GW signal is still discernible from the detector noise and the efficiency is not strongly affected by noise fluctuations.

\subsection{Gravitational wave searches with neutrino trigger}
Figure~\ref{Fig. Eff TimeFreq and cWB} shows the detection efficiency for our proposed methods, assuming a source distance of $0.5$~kpc and a false alarm rate of one per year. The top panel of the figure displays the efficiency of the MF method which only uses  the SASI timing information to construct the filters. 
The middle panel shows the detection efficiency of the MF methods with  filters  constructed by first selecting the SASI time window(s) and  applying a bandpass filter.
The bottom panel compares the detection efficiency of two selected MF methods with the 
efficiency of the two EP searches.
All our MF methods achieve an efficiency greater than $90\%$ for all three waveforms, outperforming the
EP searches by approximately $30\%$.

The methods considering both SASI periods into account, MF($t_{1,2}$) and MF($t_{1,2},f$), perform worse
than the filters built from a single SASI period, MF($t_{1}$), MF($t_{2}$), MF($t_{1},f$) and MF($t_{2},f$). 
However, the average performances of the MF methods are within $\sim 5\%$ from each other.

The degradation in efficiency we observe when using information from both SASI
periods (cf.~top panel of Fig.~\ref{Fig. Eff TimeFreq and cWB}) can be understood based on the underlying physics. 
For our s27 model,  the first SASI episode shows a predominant sloshing mode (i.e., SASI tends to develop along a particular plane, despite its global nature), however as SASI becomes stronger, it acquires a spiral nature~\cite{Hanke:2013ena}. This implies that the observer direction pointing towards the strongest SASI modulations evolves with time.

The transverse nature of GWs means that 
the strongest signal will be seen by observers situated perpendicular to the plane in which the 
SASI activity and therefore the neutrino signal modulations is strongest. This implies that the observer directions of strongest, intermediate and weak SASI modulations selected for the neutrino event rate by relying on the first SASI episode are not anymore optimal for the second SASI episode.
Additionally, in our benchmark CCSN model, the two SASI periods do not spatially align. Therefore, the chance of selecting observer directions that present an
optimal correlation between the neutrino and GW signals in both SASI windows
is low. This implies that, using both SASI periods typically leads to a reduction in the total SNR and a
reduction in detection efficiency. 

An additional uncertainty linked to the neutrino signal is the flavor mixing.
Because of the uncertainties linked to the modeling of the flavor conversion of neutrinos in dense media~\cite{Tamborra:2020cul,Mirizzi:2015eza,Richers:2022zug,Duan:2010bg}, we here consider two extreme scenarios: absence of flavor mixing (i.e., we consider the $\bar\nu_e$ signal) or maximal flavor mixing (i.e., we consider the $\bar\nu_x$ signal). 
This choice is motivated by the fact that the SASI signatures in the neutrino signal are flavor dependent, see e.g.~Refs.~\cite{tamborra_13,tamborra_14b} for a dedicated analysis of the neutrino signal linked to our benchmark supernova model.

We select only two scenarios of Tab \ref{Tab: TimeFreq selections} for which we consider full flavor conversion: the ones focusing on the first time period of SASI, with and without the frequency selection (i.e., MF($t_1$) and MF($t_1,f$)). The corresponding efficiencies are shown in Fig.~\ref{Fig. Eff TimeFreq and cWB} (blue dashed lines). We can see that the efficiencies are slightly worse than the results without flavor conversion in both cases, showing a variance of no more than $2\%$.

\subsection{Comparison with gravitational wave unmodeled search}

\begin{figure}
 \centering
 \includegraphics[width=0.5\textwidth]{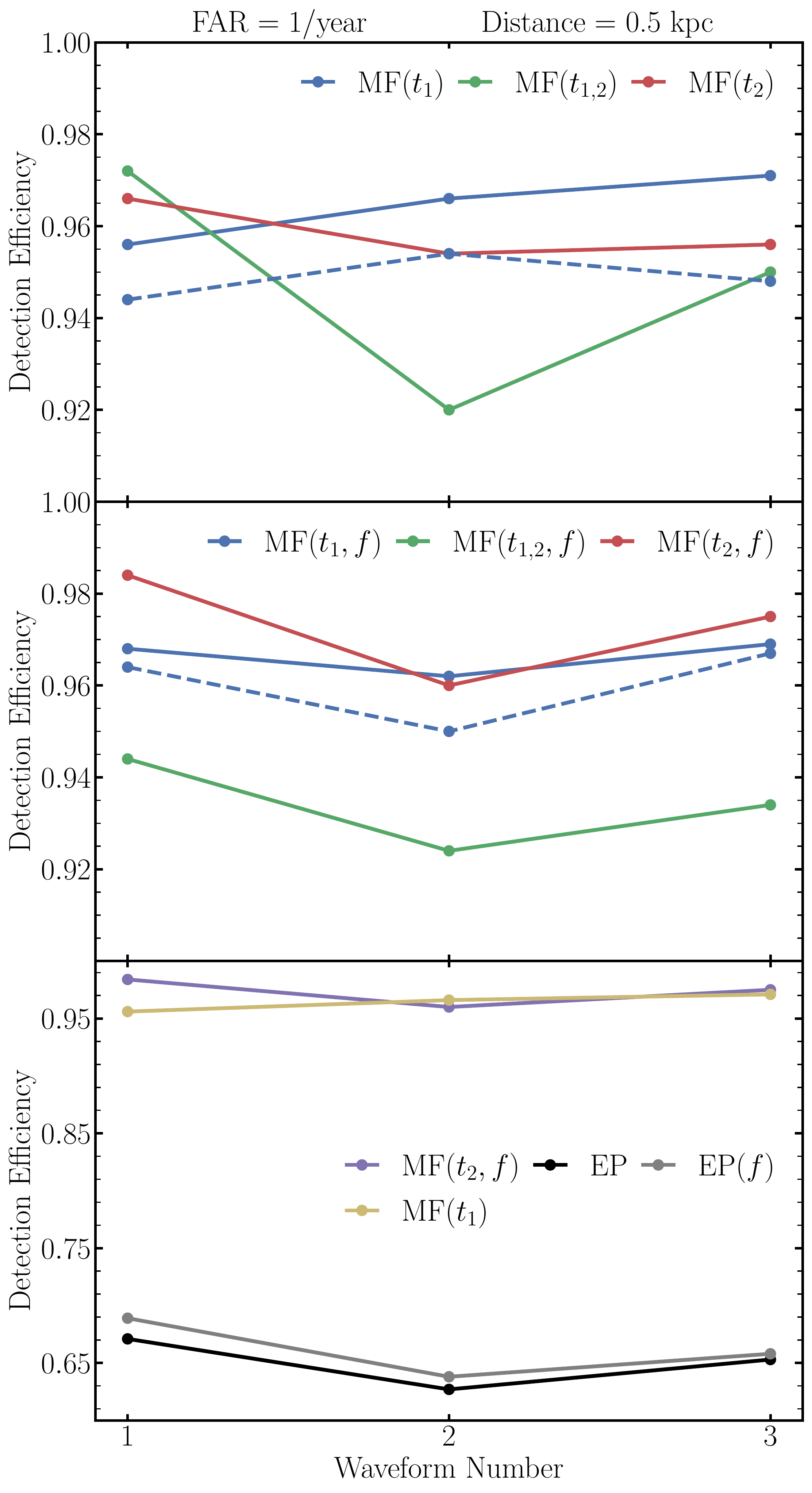}
 \caption{Detection efficiency for our three selected waveforms and
 different detection strategies. The top panel shows the MF method where 
 the filter was constructed by relying on the SASI time periods reconstructed from
 neutrino signal. The middle panel shows the MF method, with a
 band pass filter.
 The bottom panel compares the best performing MF curves of the top and middle panel
 to the standard EP method. The black curve represents the fiducial cWB search,
 while the gray curve represents a cWB search where we have restricted the
 frequency range to be $[50,300]$~Hz. 
 The solid lines refer to the scenario with no flavor mixing, while the  dashed lines represent
the case of full flavor conversion. 
 All efficiencies are calculated at a $FAR \approx 1/$year and a distance of $0.5$~kpc. Note that the y-axis in the bottom panel differs from the other two.}
 \label{Fig. Eff TimeFreq and cWB}
\end{figure}

Figure~\ref{Fig. Eff TimeFreq and cWB} displays the efficiency achieved by the unmodeled GW search method in comparison to
our proposed methods. We observe that the efficiency of the unmodeled search is lower than the other methods,
by up to $30\%$. However, we note that the parameters for the unmodeled search were taken from the official LIGO-Virgo 
analysis during the second scientific run \cite{LIGOScientific:2019ryq}, which only considered two GW detectors. 
As our analysis includes data from three detectors, the running settings are not optimized for our case, which explains 
why the efficiency in Fig.~\ref{Fig. Eff TimeFreq and cWB} does not reach $100\%$ even at a small distance ($ <0.5$~kpc).
We do not fine-tune these parameters in this work as it is beyond the scope of our study. 
Moreover, recent developments of the cWB algorithm \cite{Klimenko:2022nji} 
may further impact the results in the future.

\subsection{Supernova detection horizon}
\begin{figure}[t!]
 \centering
 \includegraphics[width=0.5\textwidth]{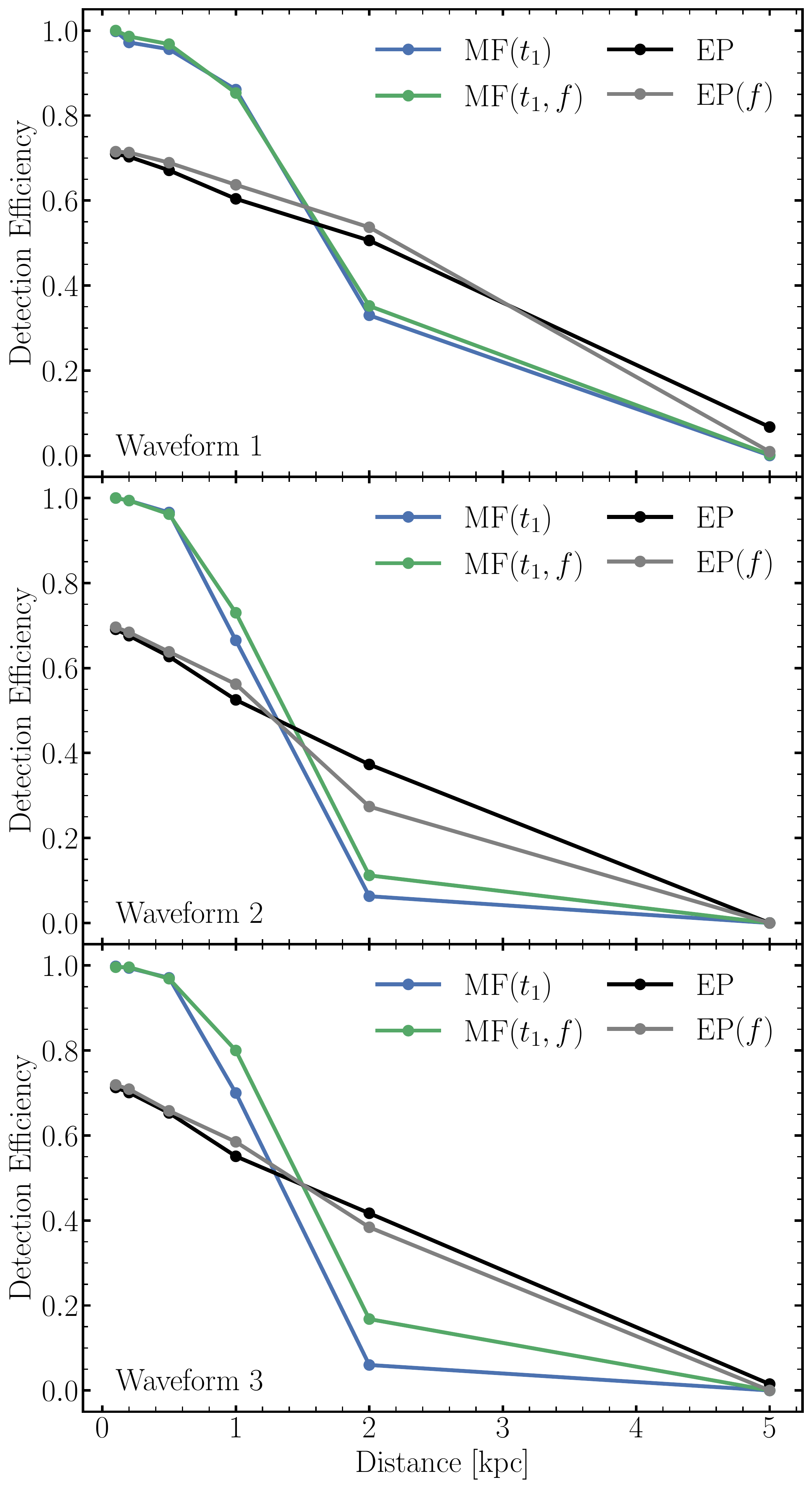}
 \caption{Detection efficiency as a function of the CCSN distance for the MF and EP methods.
 The top, middle, and bottom panels represent the detection efficiency for waveform 1, 2, and 3, respectively. For CCSNe occurring beyond $1.5$~kpc, the efficiency of the MF methods rapidly degrades, and the EP approach tends to perform better.}
 \label{Fig. Horizon Comparison}
\end{figure}

We now assess the detection horizon of our methods. 
We assume CCSNe occurring at distances of $0.1$, $0.2$, $0.5$, $1$, $2$, and $5$~kpc and calculate the efficiency as a function of the CCSN distance.

The corresponding efficiency curves as a function of the CCSN distance are reported in Fig.~\ref{Fig. Horizon Comparison}. We can see that $50\%$ of the efficiency is obtained at $2$ kpc. At smaller distances, any input from the neutrino event rate in the MF method allows to achieve an efficiency larger than the one obtained for the EP search. However, at larger distances the efficiency of the MF methods rapidly degrades, and the
EP approach tends to perform better.

The worse performance of the MF methods for larger CCSN distances 
could be due to the fact that the neutrino and the GW signals are not exactly matching.  Therefore, when the SNR decreases, the noise hinders the matched filter efficiency.

\section{Discussion}
\label{sec_discussion}

We have explored the potential of using the SASI imprints in the neutrino signal to improve the GW detection techniques. Our results are encouraging, but further improvements are required. For example, the GW detection efficiency would further improve by applying advanced methods for constructing the filter from the neutrino signal. 
One possible way forward consists in developing a set of filters that permit a time-dependent phase difference between the filter and the underlying neutrino signal. Such an approach could potentially account for the differences in the time evolution of the GW and neutrino signals. Also, the use of EP methods, such as cWB, should be explored by substantially modifying the pipeline to include information from the SASI time intervals. Moreover, a recent version of the cWB pipeline was recently published \cite{Klimenko:2022nji}, which can 
further improve the efficiency for CCSN detection.

Yet, our results may also be affected by the strength and duration of the SASI episodes, which depend on the CCSN model adopted as benchmark. In this sense, our model s27 should be considered as representative of an average CCSN; additional work is required to further assess the GW detection efficiency for a larger number of CCSN models with varying properties. For example, we may expect an overall improvement of the GW detection horizon, possibly exploiting the strong and long-lasting SASI imprints from black hole forming collapses~\cite{Walk:2019miz}. On the other hand, moderate rotation smears the SASI signatures on the neutrino signal~\cite{Walk:2019ier,Walk:2018gaw}, in which case the neutrinos may not improve the GW detection efficiency. 

We also stress that we focused on considering the neutrino event rate expected in the IceCube Neutrino Observatory, since this would provide the largest event rate for close CCSNe~\cite{Scholberg:2017czd,Mirizzi:2015eza}. However, similar analyses could be carried out by relying on other neutrino telescopes, such as Hyper-Kamiokande~\cite{Abe:2011ts}, which will be more suitable for CCSNe occurring at larger distances because the background will be essentially absent---see discussion in Sec.~\ref{sec_detectors} and Ref.~\cite{tamborra_13}. 
In particular, the latter approach could be especially interesting for third generation GW detectors, like Einstein Telescope, for which we expect a CCSN detection horizon improved by one order of magnitude~\cite{Maggiore:2019uih}.

The existing literature considers various GW search methods. In the following we compare our findings to existing work.
Note, however, that for a consistent comparison, it is necessary to have a similar CCSN model with comparable properties and affine working point, i.e. same detector sensitivities, interferometer network, and chosen FAR.

An optically triggered search for GWs of five CCSN events occurred in coincidence with the first and second observing LIGO-Virgo runs is carried out in Ref.~\cite{Abbott:2019pxc}.
However, this analysis considers a different FAR and detector network; the sensitivity of the instruments and the results of the search are not comparable to ours.

Reference \cite{Szczepanczyk:2021bka} explored the detectability of GWs from a Galactic CCSN using cWB and considering a two-detector LIGO network with expected sensitivity of the future fifth LIGO-Virgo-Kagra observation run (the O5 run).
While some of their CCSN models are similar to our model s27, there are severe differences between our analysis and the one presented in Ref.~\cite{Szczepanczyk:2021bka}. Importantly, the FAR and the detector network considered by \cite{Szczepanczyk:2021bka} differs from what we use in this work.
While these differences make a direct comparison difficult, our approach seems to result in an improvement of the CCSNe GW detection horizon by an order of magnitude over what was reported by \cite{Szczepanczyk:2021bka}.

A strategy to combine GW and neutrino searches through a combined global false alarm rate was developed in Ref.~\cite{Halim:2021yqa}. 
Their figures 2 and 6 display the 
efficiency curves obtained for different GW emission models at two FAR working points of $864$/day and 1/1000 years, respectively without and with combination with neutrino information.
Because of the difference in both of the GW waveforms used and the chosen FAR, it is challenging to assess whether 
our approach could lead to any further improvements over the method implemented in Ref.~\cite{Halim:2021yqa}.
Reference~\cite{Lin:2022jea} does not provide information about FAR, hence a direct comparison with our work is not possible.

The authors of Ref.~\cite{Nakamura:2016kkl} adopted likelihood approach, searching for a signal in the frequency band of $[50$-$500]$ Hz.
Their approach is similar to our EP($f$) method but relies on a different frequency band. Since
the authors of Ref.~\cite{Nakamura:2016kkl} do not report their FAR, a direct comparison with our method is not possible.

\section{Conclusions}\label{sec_conclusions}
This work has explored the feasibility of enhancing the detection efficiency of GWs from CCSNe by leveraging on the correlation between the GW and neutrino signals in the presence of the SASI hydrodynamical instability. In order to do so we rely on a benchmark full-scale three-dimensional CCSN simulation with zero-age main sequence mass of $27\ M_\odot$ and  consider two different approaches; one in which we use the neutrino event rate to build a matched filter for the GW signal (matched filter method), and another one that relies on  unmodeled GW search with a frequency bandpass inferred from the neutrino signal (excess power method). 

Our key findings are:
\begin{enumerate}
 \item{Utilizing the correlation between the neutrino and GW signals from  CCSNe, our matched filter method can increase the detection efficiency, compared
 to standard excess energy searches, by up to $30\%$ for nearby
 galactic events (closer than $1.5$~kpc). However, at distances above $1.5$~kpc we observe a decrease
 in the overall detection efficiency when incorporating input from the neutrino event rate.}

 \item{Combining information from two distinct SASI episodes, for the same CCSN event and for fixed observer direction,
 is in general a bad strategy and leads to a degradation of the detection
 efficiency. This decreased efficiency arises from an observer effect: because of the sloshing and spiral nature of SASI, the associated GW signal modulation is not 
 as strong for both periods for any given observer.} Therefore, attempting to combine 
 information from two different SASI periods can result in a decrease of the SNR, ultimately reducing the overall detection sensitivity. 
 
 \item{Incorporating neutrino information into the Excess Power 
 method does not significantly improve the detection sensitivity. 
 In fact, restricting the frequency band effectively reduces the FAR, 
 but it also decreases the SNR of the GW signal. These two effects balance each other out, resulting in no real change in the final detection sensitivity.}
\end{enumerate}

We have limited our investigation to one specific CCSN model for which the detection horizon is limited to a few kpc. However, results are qualitatively independent of the chosen model. We stress that our method is strongly related to the SASI signatures, so performances would improve for CCSN models with stronger SASI. Future work should include more CCSN models and consider more advanced methods for constructing the filters from the neutrino signal.

To conclude, multi-messenger techniques could improve the detection horizon of GWs from CCSNe. This work is a first attempt showing interesting potential of using the imprints of hydrodynamical instabilities in the neutrino signal to aid the GW detection. Additional work is required to further assess the performance of such technique and advice the best strategy for the first GW detection from CCSNe.
Further work is required to fully evaluate the efficacy of this technique and to determine the most effective strategy for achieving the first GW detection from CCSNe.

\begin{acknowledgments}

We would like to thank Tito Dal Canton for useful discussions and constructive inputs. The authors are grateful for
computational resources provided by the LIGO Laboratory and supported by National
Science Foundation Grants PHY-0757058 and PHY-0823459.
IDP and MD acknowledge the support from the Amaldi Research Center funded by the MIUR program ``Dipartimento di Eccellenza'' (CUP:B81I18001170001), the Sapienza School for Advanced Studies (SSAS) and the support of the Sapienza grants: RG12117A87956C66 and RM120172AEF49A82.
The work of ATF was funded by the Spanish Agencia Estatal de Investigaci{\'o}n (Grants No.~PGC2018-095984-B-I00 and PID2021-125485NB-C21) funded by MCIN/AEI/10.13039/501100011033 and ERDF A way of making Europe and by the Generalitat Valenciana (PROMETEO/2019/071). IT thanks the Villum Foundation (Project No.~37358), the Danmarks Frie Forskningsfonds (Project No.~8049-00038B), and the Deutsche Forschungsgemeinschaft through Sonderforschungbereich SFB 1258 ``Neutrinos and Dark Matter in Astro- and Particle Physics'' (NDM). HA was supported by the Swedish Research Council (Project No.~2020-00452)
\end{acknowledgments}

\bibliography{refs}

\end{document}